\documentstyle[preprint,aps]{revtex}
\begin{document}
\draft
\preprint{}
\title{A General Relativistic Model for Confinement in SU(2)
Yang-Mills Theory}
\author{D. Singleton and A. Yoshida}
\address{Department of Physics, University of Virginia,
Charlottesville, VA 22901}
\date{\today}
\maketitle
\begin{abstract}
In this paper we present a model of confinement based on an analogy
with the confinement mechanism of the Schwarzschild solution of
general relativity. Using recently discovered exact, Schwarzschild-like
solutions of the SU(2) Yang-Mills-Higgs equations we study the behaviour
of a scalar, SU(2) charged test particle placed in the gauge fields of
this solution. We find that this test particle is indeed confined
inside the color event horizon of our solution. Additionally it is
found that this system is a composite fermion even though there are
no fundamental fermions in the original Lagrangian.
\end{abstract}
\newpage
\narrowtext
\section{Introduction}
Recently a new classical solution was discovered
by one of the authors \cite{sing}
for an SU(2) Yang-Mills-Higgs system, which could be considered
the Yang-Mills version of the Schwarzschild solution of
general relativity. By using the Wu-Yang ansatz \cite{wu}
\begin{eqnarray}
\label{wu}
W_i ^a &=& \epsilon _{aij} {r^j \over g r^2} [1-K(r)] \nonumber \\
W_0 ^a &=& {r^a \over g r^2} J(r) \nonumber \\
\phi ^a &=& {r^a \over g r^2} H(r)
\end{eqnarray}
in the Euler-Lagrange equations for an SU(2) gauge field ({\it i.e.}
$W_{\mu} ^a$) coupled to a triplet scalar field ({\it i.e.} $\phi ^a$)
in the BPS limit \cite{bogo} \cite{prasad}, a solution of the following
form was found
\begin{eqnarray}
\label{singsol}
K(r) &=& {C r \over 1 - C r} \nonumber \\
J(r) &=& {B \over 1 - C r} \nonumber \\
H(r) &=& {A \over 1 - C r}
\end{eqnarray}
Similiar solutions to pure Yang-Mills theory were discovered
separately by Lunev \cite{lunev}, Mahajan and Valanju \cite{maha},
and Swank {\it et. al.} \cite{swank}.
The constants $A$ and $B$ must satisfy $A^2 - B^2 = 1$, and the
constant $C$ sets the size scale of the solution. The gauge and
scalar fields of this solution develop singularities at
$r=0$ and also on a spherical shell $r = r_0 = 1/C $.
This is also what happens with the Schwarzschild solution of
general relativity in Schwarzschild coordinates. (For the
Schwarzschild solution however the singularity on the spherical
shell is an artifact of the coordinates which are used, as can be
seen by using Kruskal coordinates.
The singularities of our solution are true
singularities, and can be thought of as the locations of the
color charge of the system, in the same way that the singularity
of the Coulomb potential of electromagnetism is the location
of electric charge). The similiarity between the Schwarzschild
solution and our solution can further be seen by comparing the
connection coefficients of the Schwarzschild solution with the
gauge fields from Eqs. (\ref{wu}) (\ref{singsol}).

What is the physical significance of these new solutions ? Aside
from indicating that there may be some deeper connection between
general relativity and non-Abelian gauge theories
\cite{utiyama} \cite{lunev2}, it was speculated that these field
configurations may give a confinement mechanism for Yang-Mills
theories in general, and QCD in particular. There were several
arguments given in Ref. \cite{sing} as to why one gets confinement
from these solutions. First, in analogy  with the Schwarzschild
solution, which permanently traps any particle carrying gravitational
charge ({\it i.e.} mass-energy) that crosses the event horizon, our
solutions would classically trap any test particle with color
charge which went inside $r_0 = 1/C$ (since the SU(2) gauge theory
we are considering is a vector theory one can have repulsion as well
as attraction, so there would be cases where a test particle would
be permanently excluded from the region $r \le 1/C$). Second, our
solution has a structure similiar to phenomenological bag models
which are used to investigate hadron dynamics. (For a review of
bag models see Ref. \cite{bag}). Finally, Ref. \cite{quigg}
reviews an argument for color confinement where the
QCD vacuum is treated as analogous to
a near perfect dia-electric medium, with
$\epsilon _{medium} \ll 1$. In this scenario a spherical
hole is postulated to exist inside the dia-electric vacuum,
and a charge is placed at the center of the hole. This
central charge will induce a charge on the surface of the
spherical hole. Since the dielectric constant is taken
as less than unity the induced charge has the same
sign as the central charge, which makes this configuration
stable against collapse (unlike the usual case where
$\epsilon_{medium} >1$ so that the induced charge and the
central charge have opposite signs).
This is similiar to the configuration of our solution, if
one takes the singularities of our solution to indicate
the locations of color charge.

In this paper we would like to examine the question of whether
this solution displays confinement in some more detail. In analogy
with the hydrogen atom bound state system, we will treat the solution
of Eqs. (\ref{wu}) (\ref{singsol}) as a spinless ``particle''
which produces the Schwarzschild-like
Yang-Mills field configuration around it. This is the interpretation
given for a similiar singular solution in Ref. \cite{swank}. In this
way the ``particle'' is taken to be located at the field singularities
(in the same way that the proton is taken to be located at the
singularity of the Coulomb potential) while the gauge fields are
taken as a background potential whose energy is not included in the
problem. The reason for taking this approach is that the field
energy of the Schwarzschild-like solution is infinite because
of the singularities. In the case of other known
solutions of the SU(2) Yang-Mills-Higgs system ({\it e.g.}
the 't Hooft-Polyakov monopole \cite{thooft} and the BPS dyon
\cite{bogo} \cite{prasad}) this problem does not arise, since these
solutions have finite energy, and thus can be taken as particles
in a straight forward way. However, it can be shown \cite{swank}
that it is impossible for these finite energy
solutions to form bound states with a test charge.
For the Schwarzschild-like solutions we will find that
not only is the test particle bound by the background potential
field, but it is permanently confined inside the sphere
$r= 1/C$. Throughout this paper we will ignore any quantum
corrections to the classical solutions. Swank {\it et. al.}
\cite{swank} have pointed out that $q {\bar q}$ creation near
$r = 1/C$ would tend to decrease the strength of the barrier
presented by the singularity. Mahajan and Valanju \cite{maha}
\cite{maha2} smooth the singularity on
the sphere by giving a phenomenological
parameterization of the quantum effects, and claim that in many
cases there is still a substantial barrier at $r=1/C$. Without
a full quantum treatment of these solutions it is not possible
to give any definite answer of how much the quantum effects
will alter the nature of the classical solutions. There are some
known methods \cite{tomb} for quantizing the finite energy monopole
and dyon solutions, which may be applicable to the Schwarzschild-like
solutions. For simplicity we will take
time component of the gauge field, $W_0 ^a$,
to be zero by chosing $B=0$ so that $A=1$, and we will take our test
particle to be spin-0. Then in order to obtain the motion of the test
particle in the gauge potentials, we study the Klein-Gordon
equation of the particle, minimally coupled to the fields of
Eqs. (\ref{wu}), (\ref{singsol}). The resulting equation can be
reduced to a one dimensional Schr{\" o}dinger-like equation.
Solving this equation shows that the test particle does indeed
remain trapped inside the region $r \le 1/C$. In the process of
reducing the minimally coupled Klein-Gordon equation to a one
dimensional Schr{\"o}dinger-like
equation we will find that the total
angular momentum of our system  equals the usual orbital angular
momentum plus the isospin of the test particle. Therefore this
system can have a spin $1/2$ even though it contains only
bosons. This effect is just the spin from isospin mechanism
discussed by Jackiw, Rebbi,'t Hooft and Hasenfrantz
\cite{hasen}. In Refs. \cite{hasen} it was shown
that bringing a particle with isospin $1/2$ into the presence of
a 't Hooft-Polyakov monopole resulted in the total system having
the spin of a fermion. That the same thing happens in our case
should not be a surprise, since our solution has the same
mixing of spatial and group indices -- see Eq. (\ref{wu}) --
that resulted in the spin $1/2$ in Refs. \cite{hasen}.
One advantage of our composite system over those considered
in Refs. \cite{hasen} is that the Schwarzschild-like field
configuration provides its own binding mechanism. In the case
of an isospinor particle moving in the field of a 't Hooft-Polyakov
monopole one has to postulate some additional,
phenomenological binding force to bind the isospinor and
monopole together. It has also been shown \cite{gold} that
such composite spin 1/2 systems obey Fermi-Dirac statistics.
Thus the bound state system that we consider in this paper,
consisting of a scalar particle with color charge moving in the
potential of the Schwarzschild-like solution,
is actually a composite fermion even though the original
Lagrangian contains only bosonic fields.

\section{Quantum Motion of Scalar, SU(2) Charged Test Particle}

The motion of a scalar particle carrying SU(2) charge,
in presence of gauge fields $W_i ^a$ (where $W_0 ^a = 0$)
is given by the minimally coupled Klein-Gordon equation
\begin{equation}
\label{kge}
\Big( \partial _i -{i \over 2} g \sigma ^a W^a _i \Big)
\Big(\partial _i - {i \over 2} g \sigma ^a W^a _i \Big) ^A _B
\Phi ^B (x) = - (E^2-m^2) \Phi ^A (x)
\end{equation}
where the scalar field was taken to have the usual time
dependence - $\Phi ^A (x,t) = e^{-iEt} \Phi ^A (x)$.
The scalar particle $\Phi ^A (x)$ is in the fundamental
representation of SU(2)  and has a mass $m$. The matrices
$(\sigma ^a)^A_B$ are the Pauli matrices with $a =1,2,3$.
and $A,B = 1,2$. $E$ is the total energy of the scalar particle.
Substituing the Wu-Yang ansatz for the gauge fields into this
gives
\begin{equation}
\label{kge2}
\left (\nabla ^2 - {[1 - K(r)] \over r^2} \sigma ^a l^a -
{[1 - K(r)]^2 \over 2 r^2} \right) ^A _B \Phi ^B (x) = -(E^2
-m^2) \Phi ^A (x)
\end{equation}
where $l_a = - i \epsilon _{aij} r^i \partial ^j$ is the orbital
angular momentum operator. In order to deal with the $\sigma ^a
l^a$ term we define the following operator
\begin{equation}
\label{isospin}
J^a = l^a + {1 \over 2} \sigma ^a
\end{equation}
This operator, which is a combination of the orbital angular momentum
operator and the isospin operator of the particle $\Phi ^A (x)$, is
in fact the total angular momentum of the system, since
it commutes with the Hamiltonian of the system.
In addition $J^a$ obeys the usual commutation relationships of
total angular momentum -- $[J^a , J^b] = i \epsilon ^{abc} J^c$
and $[J^a , P^b] = i \epsilon ^{abc} P^c$, where $P^a$ is the
canonical momentum. A more thorough demonstration that $J^a$ is
indeed the total angular momentum of the system is given in
Ref. \cite{hasen}.

Eq. (\ref{kge2}) can be reduced to a one dimensional
Schr{\"o}dinger-like equation by taking $\Phi ^A = {1 \over r}
f_{Jl} (r) Y_{JlM} ^A (\theta , \phi)$, where
$f_{Jl} (r)$ is a radial function, and
$Y_{JlM} ^A (\theta , \phi)$ are the spherical harmonics
associated with the operator $J^a$ of Eq. (\ref{isospin}). These
spherical harmonics obey the usual operator eigenvalue equations --
$J_{op} ^2 Y_{JlM} ^A = J(J+1) Y_{JlM} ^A$ and
$l_{op} ^2 Y_{JlM} ^A = l(l+1)
Y_{JlM} ^A$. Using all this in Eq. (\ref{kge2}) yields
\begin{eqnarray}
\label{kge3}
\Bigg( {d^2 \over d r^2} - {l(l+1) \over r^2} +
{J(J+1) - l(l+1) - {3 \over 4} \over r^2} [1- K(r)] &-&
{[1-K(r)]^2 \over 2 r^2} \Bigg) f_{Jl} (r) =
\nonumber \\
&-& (E^2 -m^2) f_{Jl} (r)
\end{eqnarray}
Using $K(r) = C r / (1 - Cr)$, making a change of variables
to $x = Cr$ and collecting terms yields the final form
of the Schr{\"o}dinger-like equation which we wish to solve.
\begin{equation}
\label{kge4}
\left( {d^2 \over d x^2} - {E x^2 + F x + G \over x^2 (1-x)^2}
\right) f_{Jl} (x) = - {D \over C^2} f_{Jl} (x)
\end{equation}
where
\begin{eqnarray}
D &=& E^2 - m^2 \nonumber \\
E &=& -2 J(J+1) + 3 l (l+1) + 7/2 \nonumber \\
F &=& 3 J(J+1) - 5l(l+1) - 17/4 \nonumber \\
G &=& -J(J+1) + 2l(l+1) + 5/4
\end{eqnarray}
Eq. (\ref{kge4}) is a simple second order differential equation,
which looks like a one dimensional Schr{\"o}dinger equation
with the first term on the left hand side as the kinetic term, and
the second term on the left as the potential term. The term on the
right hand side acts as the energy eigenvalue. A typical example
of the form of the ``potential'' is given in Fig. (1), which shows
the $l=0 , J=1/2$ and $l=1 , J=1/2$ cases. From this,
one can see that the particle will remain trapped between the two
barriers at $x=0$ and $x=1$. This problem is similiar
to the P{\"o}schl-Teller potential hole \cite{flugge}, which also
confines a particle between two singularities that are of the same
order as those in our problem. Unfortunately, when one attempts
to find a power series solution for Eq. (\ref{kge4}) an extremely
complicated recursion relationship occurs which makes it difficult
to get much physical insight from the analytic solution.
This makes it more convenient (without sacrificing any physical
insight) to obtain the solutions numerically.
Figs. (2) and (3) show the radial functions,
$f_{Jl} (x)$, of the first two energy levels for $J= 1/2$,
$l=0$ and $J=1/2$, $l=1$ respectively. From these figures one can
see that the scalar particle is confined between the two singularites
at $x=0$ and $x=1$. The radial function $f_{Jl}$ (and therefore
$\Phi ^A$) becomes zero for $r > r_0$. Since we solved the one
dimensional Schr{\"o}dinger equation numerically we needed some
method for taking the singularity at $x=1$ into account.
This was accomplished by requiring that $f_{Jl} (x)$ be zero
if one of our mesh points was on the singularity. This
condition is exactly the same as in ordinary quantum mechanics where
the wavefunction is required to vanish where ever the potential
becomes infinite. Pinning $f_{Jl}$ to be zero at $x=1$ and
solving Eq. (\ref{kge4}) for the range $0 \le x \le 2$ results
in the particle being entirely contained in the region $x < 1$.

In Table (I) we list the first four energy levels for the
``potentials'' that results by setting $l = 0,1,2,3 $. The
eigenvalues for the $l=0$, $J=1/2$ state look similiar
to the energy spectrum for the one dimensional
nonrelativistic infinite square well
({\it i.e} the first excited state is approximately 4 times the
ground state, the second excited state is approximately 9
times the ground state, {\it etc.}) except that the eigenvalues
are the square of the energy, $E^2$, rather than the energy,
$E$. This is not surprising since
the $l=0 , J = 1/2$ ``potential'' in Fig. (1) is a reasonable
approximation of a square well, and Eq. (\ref{kge4}) is similiar
to the one-dimensional Schr{\"o}dinger equation. The spectrum of
eigenvalues given in Table (I) agrees roughly with those
obtained by Mahajan and Valanju \cite{maha} who studied the
motion of fermions in a similiar, SU(3) background field.

Since the composite system of the particle $\Phi ^A$ moving in
the fields of the Schwarzschild-like solution behaves as a
fermion one could try to construct a model for baryons
from this system. These toy model SU(2) baryons are bound
states of two scalar particles (the test particle, $\Phi ^A$,
and the Schwarzschild-like solution considered as a ``particle'')
whose internal angular momentum comes from the spin from isospin
mechanism. This is to be compared to the usual picture of bayrons
as bound states of three fundamental fermions
({\it i.e.} quarks). In either
case one can not seperate the constituents of the bound state.
Although a more realistic model should use SU(3) as the
gauge group rather than SU(2), the present development should
give a good qualitative idea of the structure of these
scalar-scalar bound state baryons.

First, assuming that most of the mass of the real baryons
comes from binding energy, we will take the
mass of the test particle, $\Phi ^A$, to be small
so that $D = E^2 - m^2 \approx E^2$. In order to get a numerical
value for the energy eigenvalues from Table (I), we must chose
a value for $C$, which is equivalent to chosing a radius for
spherical shell singularity of the gauge fields. Taking this
radius to be 1 fermi leads to $C \approx 200$ MeV in our units.
For the first four states of the $l=0$, $J=1/2$
system we find $E_0 = 716$ MeV, $E_1 = 1394$ MeV, $E_2 =2044$ MeV,
and $E_3 = 2685$ MeV.  Similiarly for the first four states of the
$l=1$, $J=1/2$ system we find $E_0 = 920$ MeV, $E_1 =1661$
MeV, $E_2 = 2341$ MeV, and $E_3 = 3000$ MeV. Since the energy
scales with $C$ as $E = C \sqrt{N}$ (where $N$ is the numerical
value of the eigenvalue, $D/C^2 \approx E^2 / C^2$, given in
Table (I)) one could increase (decrease) these energies by
decreasing (increasing) the radius of the spherical
singularity. In order to calculate the mass  of this
composite system it would be necessary to add the
constituent mass of
the Schwarzschild ``particle'' to the binding energy. As
mentioned in the introduction, we are treating the gauge
fields of Eqs. (\ref{wu}) (\ref{singsol}) as background
potentials, since including their energy in the
mass of the Schwarzschild ``particle'' would lead to
an infinite mass. What one can do is to
take the constituent mass of the Schwarzschild ``particle''
as a parameter, which is fixed by the measured mass of
the lowest mass state. In this way one can not calculate the
mass of the lowest state from first principles, but the
masses of all the other excited states can be calculated.
In a more detailed scheme one could also take $C$ and the
mass of the test particle, $m$, as free parameters which are
fixed using the first few states as inputs.
This may seem a somewhat shady procedure, but it is
not too different from what is done for other bound state
systems such as the hydrogen atom. In the case of the
hydrogen atom one calculates the motion of the electron in
the Coulomb potential produced by the proton. The mass of
the hydrogen atom is then found by adding the mass of the
proton, the mass of the electron and the binding energy, in
the approximation of taking the proton to be a point charge.
The energy of the Coulomb field due to the proton (which
would give an infinite contribution to the mass of the
system if treated classically) is in effect normalized
into the mass of the proton.
It is worth noting that if one wanted to identify
the 716 MeV energy state of the above baryons with the proton
most of the mass would be coming from the binding
energy. The mass parameter of the Schwarzschild ``particle''
would then be $\approx 222$ MeV. This justifies
{\it a posteriori} having the nearly massless
test particle, $\Phi ^A$, move
in the stationary field of the Schwarzschild ``particle''.
This is in qualitative agreement with the quark model
picture of the proton or neutron, where the quarks are given
a small current mass, and most of the mass is attributed to
the QCD binding.

Although this toy model of the baryons has some interesting
features (a general relativistic explanation of confinement,
and having the spin come from the isospin of the test particle)
there are many problems and questions which must first be
addressed before one could make a comaparision with real
bayrons. First one should use SU(3) rather than SU(2)
as the gauge group. We have recently discovered
the SU(N) generalization for the Schwarzschild-like
solution \cite{sing2}, so it should be possible to carry
through the development here for SU(3). The SU(N) solution is
simply an embedding of SU(2) into SU(N), so for SU(3)
we would expect results which, even numerically, are not too
different from those obtained in this paper. Comparing our
eigenvalues from Table (I) with corressponding ones calculated
by Mahajan and Valanju \cite{maha}, for fermions moving in a
similiar SU(3) potential, we find that these eigenvalues are not
drastically different. Second, these bound states do not carry any
electric charge, but do carry a topological magnetic charge due
to the scalar field $\phi ^a$ . The lack of an electric charge makes
these states bad models for charged baryons such as the proton, while
the presence of the magnetic charge makes them a bad model for any
baryons. In order to give these
bound states electric charge one could easily give the test particle
an Abelian electric charge in addition to its SU(2) color charge.
This added electric charge on the test particle should not change
the nature of the composite system much, since the electric coupling
is a small perturbation to the color SU(2) coupling.
This still leaves the topological magnetic charge coming from the
scalar field $\phi ^a$. There is a possible resolution to this.
Instead of adding an Abelian electric charge by hand
one could work with a dual non-Abelian theory, and then, according
to the conjecture of Montonen and Olive \cite{olive}, the topological
charge of the solution becomes electric rather than magnetic.
In this way the field configuration of our solution carries the
electric charge, so there is no need to have the test particle
carry the electric charge. Finally, there is the already mentioned
problem of the mass of the Schwarzschild ``particle''. If we
equate the mass with the volume integral of the energy density we
get an infinite mass from the singularities in the solution.
Another possible resolution, aside from normalizing this energy
into the constituent mass of the
Schwarzschild ``particle'', is to try and
smooth out the singularities of the solution, while still maintaining
the spherical barrier feature that leads to confinement. This can be
done by hand by allowing $1/C$ to be complex \cite{maha}. A less
{\it ad hoc} approach would be to allow the scalar field $\phi ^a$
to have a mass and/or self coupling term in the hope that the solution
with these terms present would be smooth. This is what happens in
the case of the BPS monopole as compared to the 't Hooft-Polyakov
monopole -- by letting the scalar field have a mass and a self
coupling the singularity at $r=0$ of the BPS monopole gets
smoothed out in the 't Hooft-Polyakov monopole. However when
the scalar field is allowed to have a mass and self coupling
term one must look for solutions numerically.

\section{Discussion and Conclusions}

Using recently discovered solutions
to the SU(2) Yang-Mills-Higgs system we studied the behaviour
of a scalar, SU(2) charged test
particle in the background potential of
these solutions. The main goal of this paper was to show that
these Schwarzschild-like solutions did exhibit confinement in that
they kept the test particle restricted to the region $r \le 1/C$.
By minimally coupling the Klein-Gordon equation of the test
particle, $\Phi ^A$, to the gauge fields of the solution, we
obtained a Schr{\"o}dinger-like equation whose potential term
had two infinite barriers - one at $r= 0$ and the other at
$r = 1/C$.  These barriers confined the test particle to
remain in the region $0 < r < 1/C$. In
addition to confining the test particle, the field configuration
of the Schwarzschild-like solution converted the isospin of the
test particle into real spin. This spin from isospin effect
is the same that occurs when one places a scalar SU(2)
test particle in the field configuration of a 't Hooft-Polyakov
monopole. (This is further related to the old
result in electromagnetism that the fields of an electric
charge and a magnetic monopole carry angular momentum \cite{saha}).
Goldhaber \cite{gold} has shown that such bound states, in
addition to carrying the angular momentum associated with
fermions, also obey Fermi-Dirac statistics. Therefore our
bound state system, consisting of the spinless test particle
moving inside the the field configuration of Eqs. (\ref{wu}) and
(\ref{singsol}), is a composite fermion, which results from a
theory that originally contained no spin-1/2 fields. To construct
composite, integer spin particles in the present model, one should
place spin 1/2, test particles inside the fields of the
Schwarzschild-like potential. The advantage these present composite
fermions (and bosons) have over those discussed in Refs. \cite{hasen}
is that they provide for their own binding mechanism. For the
`t Hooft-Polyakov monopole and the BPS dyon it has been demonstrated
\cite{swank} that one can not form bound states with these field
configurations. In order to get a test particle to form a bound state
with these finite energy solutions one needs to postulate some
phenomenological Yukawa binding between the two scalar particles.
In the present case the isospin 1/2, scalar particle is automatically
bound by the field configuration of our solution. Not only is it bound,
but it is permanently confined. Thus, even though the system is
a composite of the scalar test particle and the ``particle'' represented
by our solution, one can never seperate the system into its constituent
parts.

There are several possible extensions to this work. One could study the
behaviour of color charged fermions in the background field of our
solution. These states would be bosons by the same mechanism that
made the scalar particles of this paper fermions. Part of the reason
for studying scalar particles here was because of this spin from
isospin effect. (If we had worked with colored fermions the spin
from isospin effect would still have been present, however the
composite system would then have had integer spin due to the
fundamental spin 1/2 of the fermion. It is much more unusal to have
a spin 1/2 composite system built from scalars, rather than an integer
spin composite system built from fermions).
Another possibility which was not explored in this paper is to have
a phenomenological Yukawa coupling between the scalar, isovector
field, $\phi ^a$ and the scalar, isospinor field, $\Phi ^A$.
We felt that it was one of the strong points of our Schwarzschild-like
solution that it did not require such a coupling in order to bind
the isospinor particle to the Schwarzschild-like field configuration.
In this paper we looked at a special case of the
Schwarzschild-like solution, namely the case where
$W_0 ^a = 0$. It may be worthwhile to examine the more
general case, $\phi ^a \ne 0$ and $W_0 ^a \ne 0$. Finally
our Schwarzschild-like solutions were found in the BPS limit
of the field $\phi ^a$. It may be of interest to see if one
can find, even numerically, solutions with a nonzero mass and/or
self-interaction term for the scalar triplet, $\phi ^a$,
which have the confining sphere feature of Eqs. (\ref{wu}) ,
(\ref{singsol}). The hope is that allowing for a nonzero
mass and/or self-interaction may smooth out some or all
of the singularities of the solution in the same way that the
singularity in the magnetic field, at $r = 0$,
of the BPS dyon gets smoothed over in the 't Hooft-Polyakov
monopole. If this conjecture proves to be true then this
might provide a classical fix for the problems posed by
the singularities of the solution.

Our solution is completely classical. One should study the quantum
fluctuations around this classical solution. There are known
methods for doing this \cite{tomb}, and such an investigation is
currently underway.

\section{Acknowledgements} D.S. would like to thank Carolyn New
and Carolyn O'Neill for encouragement during the completion
of this work.

\newpage
\begin{center}
{\bf Figure Captions}
\end{center}

{\bf Figure 1} : The ``potentials'' of the one dimensional
Schr{\"o}dinger equation, Eq. (\ref{kge4}), for the
specific cases $l=0 , J=1/2$ and $l=1 , J= 1/2$.

{\bf Figure 2} : The radial functions, $f_{Jl} (x)$, for the
three lowest states for the case $l=0 \; J=1/2$ .

{\bf Figure 3} : The radial fucntions, $f_{Jl} (x)$, for the
three lowest states for the case $l=1 \; J =1/2$.

\begin{table}
\caption{This table gives the eigenvalues, $D_i/C^2$, to
Eq. (\ref{kge4}), for $l=0,1,2,3$.}
\begin{tabular}{c|c|clcl}
\multicolumn{1}{c}{$l$} &\multicolumn{1}{c}{$J$}
&\multicolumn{1}{c}{$D_0/C^2$} &\multicolumn{1}{c}{$D_1/C^2$}
&\multicolumn{1}{c}{$D_2/C^2$} &\multicolumn{1}{c}{$D_3/C^2$} \\
\tableline
$\; \;$  0 $\; \;$ & 1/2 &12.81 &48.57  &104.47 &180.23 \\
$\; \;$  1 $\; \;$ & 1/2 &21.16 &68.97  &136.97 &225.02 \\
$\; \;$  1 $\; \;$ & 3/2 &23.15 &64.48  &125.52 &206.31 \\
$\; \;$  2 $\; \;$ & 3/2 &31.18 &88.65  &165.89 &263.08 \\
$\; \;$  2 $\; \;$ & 5/2 &40.06 &90.57  &160.58 &250.25 \\
$\; \;$  3 $\; \;$ & 5/2 &43.37 &111.22 &198.15 &304.86 \\
$\; \;$  3 $\; \;$ & 7/2 &60.87 &121.11 &200.88 &300.18 \\
\end{tabular}
\end{table}

\end{document}